\documentclass{article}
\usepackage[table]{xcolor}
\usepackage{spconf,amsmath,graphicx,hyperref,amsfonts,amssymb, multirow}
\usepackage[marginal]{footmisc}

\makeatletter
\def\blfootnote{\xdef\@thefnmark{}\@footnotetext}
\makeatother
\title{Position-Aware Self-supervised Representation Learning for Cross-mode Radar Signal Recognition}
\name{Hongyang Zhang\textsuperscript{1,2,$\ast$}, Haitao Zhang\textsuperscript{1,$\ast$}, Yinhao Liu\textsuperscript{1}, Kunjie Lin\textsuperscript{1}, Yue Huang\textsuperscript{1}, Xinghao Ding\textsuperscript{1}}
\address{
\textsuperscript{1}Key Laboratory of Multimedia Trusted Perception and Efficient Computing,\\ Ministry of Education of China, Xiamen University, 361005, China. \\
\textsuperscript{2}School of Science and Engineering, The Chinese University of Hong Kong, Shenzhen, 518172, China.
}
\begin{document}
%
\maketitle
%
\blfootnote{\NoHyper \textsuperscript{$\ast$}Equal contribution.}
\blfootnote{\NoHyper This work was supported the National Natural Science Foundation of China under Grant Grant U19B2031.}
\blfootnote{\NoHyper In part by the Dreams Foundation of Jianghuai Advance Technology Center project under Grant 2023-ZM01D002.}

\begin{abstract}
Radar signal recognition in open electromagnetic environments is challenging due to diverse operating modes and unseen radar types. Existing methods often overlook position relations in pulse sequences, limiting their ability to capture semantic dependencies over time. We propose RadarPos, a position-aware self-supervised framework that leverages pulse-level temporal dynamics without complex augmentations or masking, providing improved position relation modeling over contrastive learning or masked reconstruction. Using this framework, we evaluate cross-mode radar signal recognition under the long-tailed setting to assess adaptability and generalization. Experimental results demonstrate enhanced discriminability and robustness, highlighting practical applicability in real-world electromagnetic environments.
\end{abstract}
\section{Introduction}
\label{sec:intro}
Radar signal recognition is a fundamental task in electronic intelligence, supporting spectrum management, situational awareness, and electronic countermeasures~\cite{r21}. It encompasses radar emitter classification, mode recognition~\cite{r13}, and the challenging discovery of unseen emitters~\cite{r14}. However, most existing methods assume a closed-world setting, where training and testing share fixed emitter types. This assumption fails in dynamic electromagnetic environments, leading to poor generalization, noise sensitivity, and limited scalability. Moreover, collecting large-scale, accurately labeled datasets for supervised learning remains highly impractical.

Real-World electromagnetic spectrum is inherently dynamic and contains unknown signal types~\cite{rouzoumka2025out}. Conventional closed-set methods, which assume all classes are known, often misclassify novel signals and suffer severe degradation~\cite{zhou2022intelligent}. This limitation has motivated research on Open-World Radar Signal Recognition, which requires classifying known signals across varying domains~\cite{lee2021radar}, while addressing domain shifts that frequently occur in real-world radar environments. Existing approaches leverage adversarial training, prototype-based constraints~\cite{liu2025open}. However, they struggle with generalization in dynamic settings and often neglect intrinsic temporal dependencies of radar pulses. To overcome these challenges, we integrate self-supervised representation learning with position-aware temporal modeling, yielding more discriminative and robust features in open environments.

\begin{figure}[!t]
\centering
\centerline{\includegraphics[width=0.9\linewidth,height=1.0\linewidth,keepaspectratio]{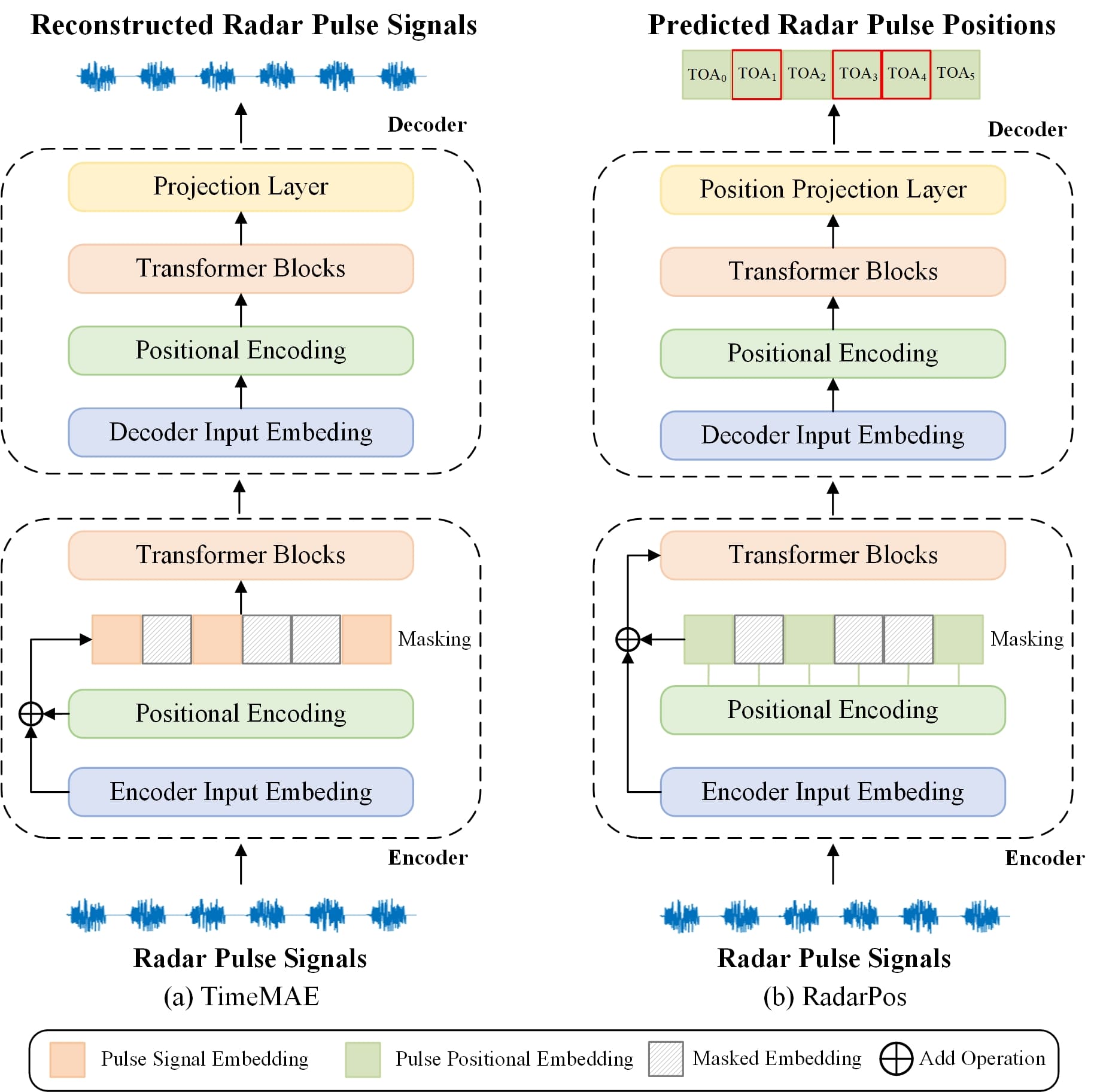}}
\vspace{-0.8em}
\caption{\small (a) MAE-style pretraining framework (TimeMAE~\cite{r9}), (b) Position-aware pretraining framework (RadarPos).}
\label{intro}
\vspace{-1.5em}
\end{figure}

Recently, more research has focused on self-supervised learning (SSL)  to extract more generalizable representations for radar signals. Nevertheless, most of them only leverage contrastive learning~\cite{r5} and masked modeling framework~\cite{r9}. They primarily exploit intra-pulse features while neglecting the correlations among inter-pulse signal sequences. Existing SSL frameworks insufficiently exploit the temporal positional dependencies of radar pulses, which are vital for discriminative radar signal feature extraction. Besides, position relation in radar also carries crucial inter-pulse modulation information like pulse repetition interval (PRI) variation or jitter, which are key discriminative cues for emitter and mode recognition~\cite{r14}. Therefore, this study emphasizes position-aware modeling for radar signal representation learning.

Based on the above analysis, we introduce position prediction as the core pretraining objective to capture inter-pulse temporal dependencies in radar signals. Without complex data augmentation or masking, the framework enhances discriminative feature learning. Moreover, Pulse Descriptor Words naturally encode time-of-arrival (TOA) information, reflecting pulse sequence positions. As shown in Fig.~\ref{intro}, existing MAE-style frameworks (e.g., TimeMAE~\cite{r9}) reconstruct masked radar pulses, focusing on intra-pulse features while neglecting temporal dependencies. In contrast, RadarPos predicts TOA, capturing inter-pulse relations and improving pulse sequence understanding. Our contributions are summarized as follows:

(1) We introduce RadarPos, a novel and efficient self-supervised pre-training framework for radar signals through position prediction, which enhances the model's ability to extract more generalizable representations.

(2) To enable efficient pre-training, we propose a position-label smoothing strategy with attention-based reconstruction to reduce errors among similar pulse signals.
  
(3) Extensive experiments on cross-mode radar signal identification downstream task demonstrate that RadarPos achieves competitive results compared with the state-of-the-art self-supervised methods.
\vspace{-10pt}

\section{Methods}
\label{sec:method}

\subsection{The overview of RadarPos Framework}
RadarPos employs position-aware pretraining via masked position prediction, reconstructing temporal patch locations from removed positional encodings. This design enhances robustness, mitigates reconstruction ambiguity, and promotes consistency between pretraining and fine-tuning, thereby improving temporal sensitivity, preventing trivial shortcuts, and strengthening both long-range dependency modeling and semantic representation. 
\begin{figure*}[!t]
\centering
\centerline{\includegraphics[width=0.9\linewidth,height=0.3\linewidth]{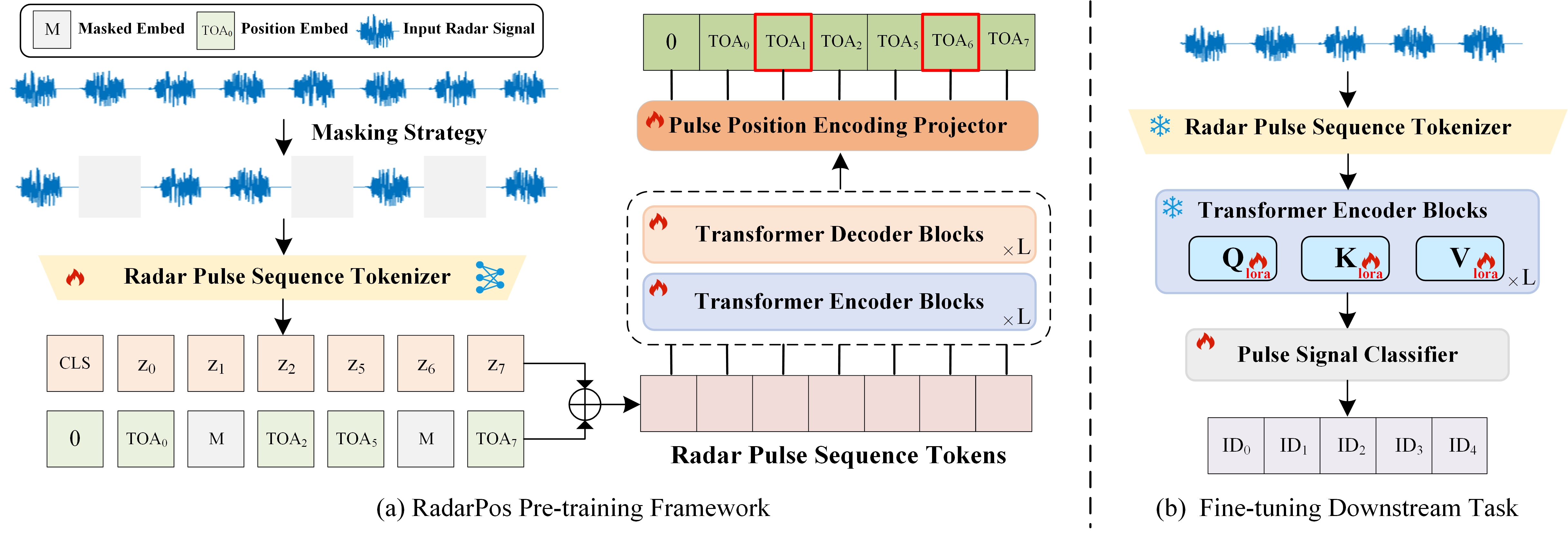}}
\vspace{-0.5em}
\caption{Overall pipeline of the RadarPos framework, (a) RadarPos Pre-training via position prediction; (b) In the fine-tuning stage, the Transformer Encoder blocks are adapted via LoRA~\cite{r18}, and the Pulse Signal Classifier is optimized.}
\label{framework}
\vspace{-1.3em}
\end{figure*}

The overall framework (Fig.~\ref{framework}) comprises four modules in the pre-training stage: (1) TOA-based positional encoding; (2) Radar pulse sequence tokenizer; (3) Transformer encoder–decoder blocks, and (4) Pulse position encoding projector. During fine-tuning, only the Transformer encoder is adapted using LoRA, while the pulse signal classifier is optimized for downstream recognition.
\vspace{-8pt}

\subsection{Position-aware Masking Strategy}

To better capture temporal dependencies in radar pulse sequences, we replace the conventional Transformer positional encoding with a TOA-based sinusoidal encoding scheme. For each radar pulse indexed by $i$, the positional embedding $p_i \in \mathbb{R}^{D}$ is computed as:
\begin{equation}
\footnotesize
p_{(i,2k)} = \sin \left( \frac{\text{TOA}_i}{10000^{\frac{2k}{D}}} \right), \quad
p_{(i,2k+1)} = \cos \left( \frac{\text{TOA}_i}{10000^{\frac{2k}{D}}} \right),
\end{equation}
where $k \in \{0,1,\dots,\frac{D}{2}-1\}$ indexes the embedding dimension. This formulation encodes the relative and absolute temporal order, allowing the model to capture long-range dependencies in radar sequences.

Furthermore, we introduce a position-aware masking strategy. Given radar pulse signals $X$ and their corresponding features $Z \in \mathbb{R}^{N \times D}$ are extracted via radar pulse sequence tokenizer, positional embeddings $p \in \mathbb{R}^{N \times D}$ 
are aligned with the sequence. A binary mask matrix 
$M_{\text{pos}} \in \{0,1\}^{N}$ with masking ratio $\alpha_{\text{pos}}$ 
is applied to randomly replace part of the positional encodings by a learnable mask token $p_{\text{mask}}$, while retaining visible positions $p_{\text{vis}}$. Overall, the masked position $p'_i$ is formulated as below:
\begin{equation}
p'_i =
\begin{cases}
p_0, & i=0, \\
p^{\,i-1}_{\text{vis}}, & \mathcal{M}^{i-1}_{\text{pos}} = 1, \\
p_{\text{mask}}, & \text{otherwise}.
\end{cases}
\end{equation}

Then, radar pulse sequence tokens are constructed as:
\begin{equation}
Z' = [z_{\text{cls}}; z_{\text{vis}}] \oplus p' \in \mathbb{R}^{(N+1) \times D}.
\end{equation}

Subsequently, these tokens are converted into transformer encoder-decoder blocks to obtain output tokens $\hat{Z}' \in \mathbb{R}^{N \times D}$. This strategy forces the model to infer masked positions, enhancing positional sensitivity and reducing ambiguity.

\subsection{Optimization on RadarPos Pre-training Stage}
A lightweight pulse encoding projector $g(\cdot)$ (A single layer MLP) is incorporated after these blocks to obtain pulse-level position logits $o = g(\hat{Z}') \in \mathbb{R}^{N \times N}$. For each position $i \in \{0, \dots, N-1\}$, the softmax $\mathrm{softmax}(o_i)$ defines the predicted probability distribution. Ground-truth masked positional labels $\mathbf{Y} \in \mathbb{R}^{N \times N}$ provide supervisory signals, with $y_{ij}$ denoting the true index for the $j$-th candidate position. The position-aware loss $\mathcal{L}_{\mathrm{pos}}$ is formulated as below:
\begin{equation}
\scriptsize
\mathcal{L}_{\mathrm{pos}} = -\sum_{i=0}^{N-1} \sum_{j=0}^{N-1} (1-\mathcal{M}_{\mathrm{pos}}^i) \cdot \mathrm{one\_hot}(y_{ij}) \cdot \log \frac{\exp(o_{ij})}{\sum_{k=0}^{N-1} \exp(o_{ik})},
\end{equation}
\noindent where $\mathcal{M}_{\mathrm{pos}}^i$ denotes mask in the $i$-th position, and $y_{ij}$ is the positional label. This objective enforces accurate reconstruction of spatial configurations and aligns predicted distributions with ground-truth positions.

In addition, we further adopt a positional smoothing strategy with attention-based reconstruction to enhance pretraining efficiency. The strategy alleviates reliance on exact positional prediction by smoothing index distributions, improving robustness to similar pulse patterns. Then, it further exploits contextual relationships to refine reconstruction and strengthen spatial reasoning.

\textbf{Positional Smoothing:}  
Since radar pulses are not fully independent, a weight matrix $\mathbf{w} \in (0,1)^{N \times N}$ is defined as
\begin{equation}
    w(i,j) = \exp \left( -\frac{\mathrm{dist}(i,j)}{\sigma^2} \right),
\end{equation}
where $\mathrm{dist}(i,j)$ is the Euclidean distance and $\sigma$ a smoothing hyperparameter, $w_i^*$ is attained by normalizing the weight matrix $w$. The loss can be defined as below:
\begin{equation}
\scriptsize
    \mathcal{L}_{\mathrm{smooth}} = -\sum_{i=0}^{N-1} \sum_{j=0}^{N-1} (1-\mathcal{M}_{\mathrm{pos}}^i)\, w^*(y_{ij},j)\, \log \frac{\exp(o_{ij})}{\sum_{k=0}^{N-1} \exp(o_{ik})}.
\end{equation}

\textbf{Attention-based Reconstruction:}  
To reduce ambiguity from similar radar pulse features, we impose feature-level similarity as an auxiliary constraint. Let $C \in \mathbb{R}^N$ denote similarity between class token $z_{\mathrm{cls}}$ and visible blocks $z_{\mathrm{vis}}$. The attention weight $C_i$ is denoted as:
\begin{equation}
    C_i = \frac{\exp(\cos(z_{\mathrm{cls}}, z^j_{vis})/T)}{\sum_{j=0}^{N-1}\exp(\cos(z_{\mathrm{cls}}, z^j_{vis})/T)},
\end{equation}
where $T$ is a temperature (the value is 0.95). The position-aware loss $\mathcal{L}_{\mathrm{smooth}}$ can be revised as below:
\begin{equation}
\scriptsize
    \mathcal{L}_{\mathrm{smooth}} = -\sum_{i=0}^{N-1} \sum_{j=0}^{N-1} (1-\mathcal{M}_{\mathrm{pos}}^i)\, C_{y_{ij}}\, w^*(y_{ij},j)\, \log \frac{\exp(o_{ij})}{\sum_{k=0}^{N-1} \exp(o_{ik})}.
\end{equation}

This design enables smoother positional estimation and uses inter-pulse correlations to reduce reconstruction ambiguity, enhancing spatial coherence of representations.


\section{Experiments}
\subsection{Implementation Details}
All experiments are conducted on an NVIDIA A40 GPU with PyTorch. Radar signals are resampled to $2 \times 512$ and min–max normalized. The transformer backbone has 6 layers, 8 attention heads, and 64 patches, with embedding dimension 512. During pre-training, AdamW is used for 300 epochs with batch size 256 and learning rate $1.0 \times 10^{-5}$. The smoothing coefficient $\sigma$ is 0.9. For fine-tuning, we adopt LoRA with rank $r=8$, optimized using AdamW for 50 epochs with an initial learning rate of $2.5 \times 10^{-5}$. The learning rate is linearly warmed up during the first 10 epochs and subsequently decayed by a factor of 0.1 every 15 epochs. The model is optimized by the cross-entropy loss and evaluated in terms of Accuracy (Acc) and F1-score (F1).

\begin{table}[ht]
\small
\vspace{-2.em}
\caption{The statistics of pre-training radar datasets.}
\vspace{0.5em}
\label{tab:pretrain_dataset}
\centering
\begin{tabular}{c|c|c}
\hline
\rowcolor{gray!15}
\textbf{Dataset} & \textbf{Number of Radar Types} & \textbf{Data Size} \\
\hline
Simulation Dataset & 300 & 3,753,200 \\
RadChar & 5 & 20,000 \\
\hline
\end{tabular}
\vspace{-2.em}
\end{table}

\begin{table}[ht]
\vspace{-1.em}
\small
\caption{The statistics of fine-tuning radar datasets.}
\vspace{0.5em}
\label{tab:longtail_dataset}
\centering
\begin{tabular}{c|c|c|c}
\hline
\rowcolor{gray!15}
\textbf{Mode} & \textbf{Training Size} & \textbf{Training Ratio} & \textbf{Test Size} \\
\hline
$m_0$ & 206 & 100:50:25:15:10:5:1 & 200 \\
$m_1$ & 206 & 100:50:25:15:10:5:1 & 200 \\
$m_2$ & 206 & 100:50:25:15:10:5:1 & 200 \\
\hline
\end{tabular}
\vspace{-2.em}
\end{table}

\subsection{The overview of radar signal datasets}
The detailed statistics of two datasets are shown in Table~\ref{tab:pretrain_dataset} and~\ref{tab:longtail_dataset}. The first part includes simulation (generated via MATLAB 2021b) and public datasets (RadChar~\cite{r16}) for pre-training, covering diverse modes and parameters. The second part is a simulation dataset designed for fine-tuning, where the training ratio reveals class distribution. It includes seven radar emitter models distinguished by three operating modes (VS, TAS, and STT), denoted as $m_0$, $m_1$, and $m_2$.
\vspace{-8pt}
\begin{table*}[!t]
\centering
\caption{Cross-mode radar signal recognition performance across three source modes: $m_0$, $m_1$, and $m_2$. The best and second-best results are highlighted in \textbf{bold} and \underline{underlined}, respectively.}
\vspace{0.5em} %
\footnotesize
\setlength{\tabcolsep}{2.0mm} 
\begin{tabular}{c|cc|cc|cc|cc|cc|cc}
\hline
\multirow{2}{*}{Method} & 
\multicolumn{2}{c|}{$m_0 \to m_1$} & 
\multicolumn{2}{c|}{$m_0 \to m_2$} & 
\multicolumn{2}{c|}{$m_1 \to m_0$} & 
\multicolumn{2}{c|}{$m_1 \to m_2$} & 
\multicolumn{2}{c|}{$m_2 \to m_0$} & 
\multicolumn{2}{c}{$m_2 \to m_1$} \\
\cline{2-13}
 & Acc & F1 & Acc & F1 & Acc & F1 & Acc & F1 & Acc & F1 & Acc & F1 \\
\hline
ResNet18~\cite{ResNet18}    & 54.50\% & 0.5264 & 55.00\% & 0.5256 & 56.50\% & 0.5154 & 59.50\% & 0.5566 & 54.00\% & 0.4500 & \underline{61.00\%} & 0.5345 \\
Transformer~\cite{Transformer} & 34.50\% & 0.2642 & 38.00\% & 0.3212 & 31.50\% & 0.2261 & 30.50\% & 0.2149 & 34.50\% & 0.2421 & 29.00\% & 0.1943 \\
IBN~\cite{IBN}         & 56.50\% & 0.5388 & 60.00\% & 0.5835 & 58.50\% & \underline{0.5530} & 60.50\% & 0.5743 & 52.50\% & 0.5009 & 60.50\% & 0.5704 \\
SNR~\cite{SNR}         & 57.00\% & 0.5426 & 61.00\% & 0.5838 & 58.00\% & 0.5325 & 57.00\% & 0.5382 & 54.50\% & 0.5177 & 60.50\% & 0.5639 \\
MixStyle~\cite{MixStyle}    & 55.50\% & 0.5133 & 61.00\% & \underline{0.5903} & 57.50\% & 0.5276 & 60.00\% & 0.5659 & 55.50\% & 0.5130 & 60.50\% & 0.5648 \\
DSU~\cite{DSU}         & 56.50\% & 0.5383 & \underline{61.50\%} & 0.5881 & 58.50\% & 0.5406 & 57.50\% & 0.5426 & \textbf{59.00\%} & \textbf{0.5582} & 60.00\% & 0.5624 \\
DIRA~\cite{DIRA}        & 56.00\% & 0.5309 & 60.50\% & 0.5722 & 56.50\% & 0.5262 & 60.50\% & 0.5521 & 57.00\% & 0.5364 & 60.50\% & 0.5613 \\
SimCLR~\cite{r5}      & \underline{58.00\%} & 0.5560 & 60.50\% & 0.5667 & \underline{60.00\%} & 0.5513 & 61.00\% & 0.5767 & 58.00\% & 0.5357 & 57.50\% & 0.5435 \\
MoCo~\cite{r6}        & \underline{58.00\%} & \underline{0.5603} & 60.50\% & 0.5860 & 59.50\% & 0.5439 & \underline{61.50\%} & \underline{0.5885} & 58.00\% & 0.5339 & \underline{61.00\%} & \underline{0.5760} \\
TimeMAE~\cite{r9}         & 50.50\% & 0.4760 & 55.50\% & 0.5216 & 50.00\% & 0.4675 & 52.50\% & 0.4930 & 55.50\% & 0.5079 & 57.50\% & 0.5346 \\ \hline
\textbf{RadarPos (Ours)} & \textbf{59.50\%} & \textbf{0.5775} & \textbf{62.00\%} & \textbf{0.5926} & \textbf{60.50\%} & \textbf{0.5591} &\textbf{62.00\%} & \textbf{0.5947} & \underline{58.50\%} & \underline{0.5570} & \textbf{62.00\%} & \textbf{0.5853} \\
\hline
\end{tabular}
\label{exp}
\vspace{-0.3em}
\end{table*}

\subsection{Quantitative Comparisons and Ablation Studies}
\subsubsection{Comparisons on cross-mode radar signal recognition}
To evaluate the performance of the proposed RadarPos framework, we conduct experiments on cross-mode long-tailed radar signal recognition in six cross-mode scenarios.

In Table ~\ref{exp}, we compare the RadarPos with recent existing SOTA methods for domain generalization and self-supervised learning, including backbones (ResNet18~\cite{ResNet18}, Transformer~\cite{Transformer}), domain generalization methods (IBN~\cite{IBN}, SNR~\cite{SNR}, MixStyle~\cite{MixStyle} and DSU~\cite{DSU}), self-supervised contrastive learning methods (DIRA~\cite{DIRA}, SimCLR~\cite{r5}, MoCo~\cite{r6} and TimeMAE~\cite{r9}).
RadarPos consistently achieves competitive results, especially in the challenging transfers $m_0\to m_1$ and $m_0\to m_2$, where it surpasses MoCo's accuracy by 1.5\% in both scenarios. This demonstrates that RadarPos not only enhances feature discriminability in common scenarios but also effectively boost cross-domain radar signal recognition in long-tailed settings.
\vspace{-12pt}
\subsubsection{Ablation studies}
We firstly conduct experiments to assess the effectiveness of TOA-based positional encoding in $m_0\to m_1$ and $m_0\to m_2$.

\begin{table}[ht]
\small
\vspace{-1.25em}
\caption{\footnotesize Ablation studies on TOA-based positional encoding (TOA PE).}
\vspace{0.5em}
\label{tab:toa}
\centering
\begin{tabular}{c|cc|cc}
\hline
\multirow{2}{*}{Method} & 
\multicolumn{2}{c|}{$m_0 \to m_1$} &
\multicolumn{2}{c}{$m_0 \to m_2$} \\
\cline{2-5}
 & Acc & F1 & Acc & F1 \\
\hline
\textbf{w/o.} TOA PE & 58.50\% & 0.5592 & 61.00\% & 0.5768 \\
\textbf{w/.} TOA PE & 59.50\% & 0.5775 & 62.00\% & 0.5926 \\
\hline
\end{tabular}
\vspace{-1.em}
\end{table}

As shown in Table~\ref{tab:toa}, incorporating TOA-based positional encoding enhances accuracy and F1 across transfer scenarios. In $m_0 \to m_1$, accuracy rises from 58.5$\%$ to 59.5$\%$, and in $m_0 \to m_2$ from 61.0$\%$ to 62.0$\%$. These findings highlight the importance of modeling inter-pulse temporal relations for more discriminative and generalizable radar recognition.

\begin{figure}[!t]
\centering
\centerline{\includegraphics[width=1.0\linewidth,height=0.4\linewidth]{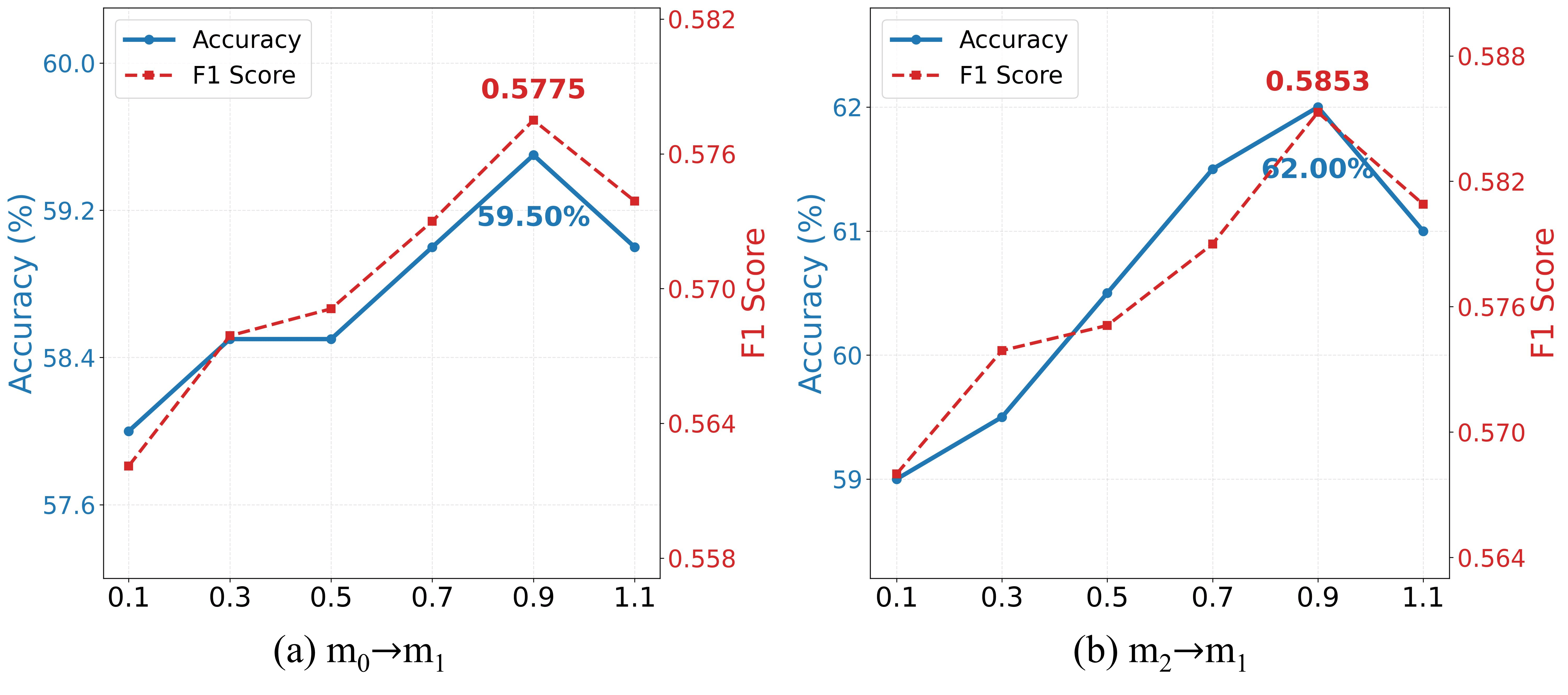}}
\vspace{-1.0em}
\caption{The sensitivity analysis of smoothing parameter $\sigma$ in two cross-mode scenarios ($m_0\to m_1$ and $m_2\to m_1$).}
\label{smooth}
\vspace{-1.0em}
\end{figure}

\begin{figure}[!t]
\centering
\centerline{\includegraphics[width=1.0\linewidth,height=0.4\linewidth]{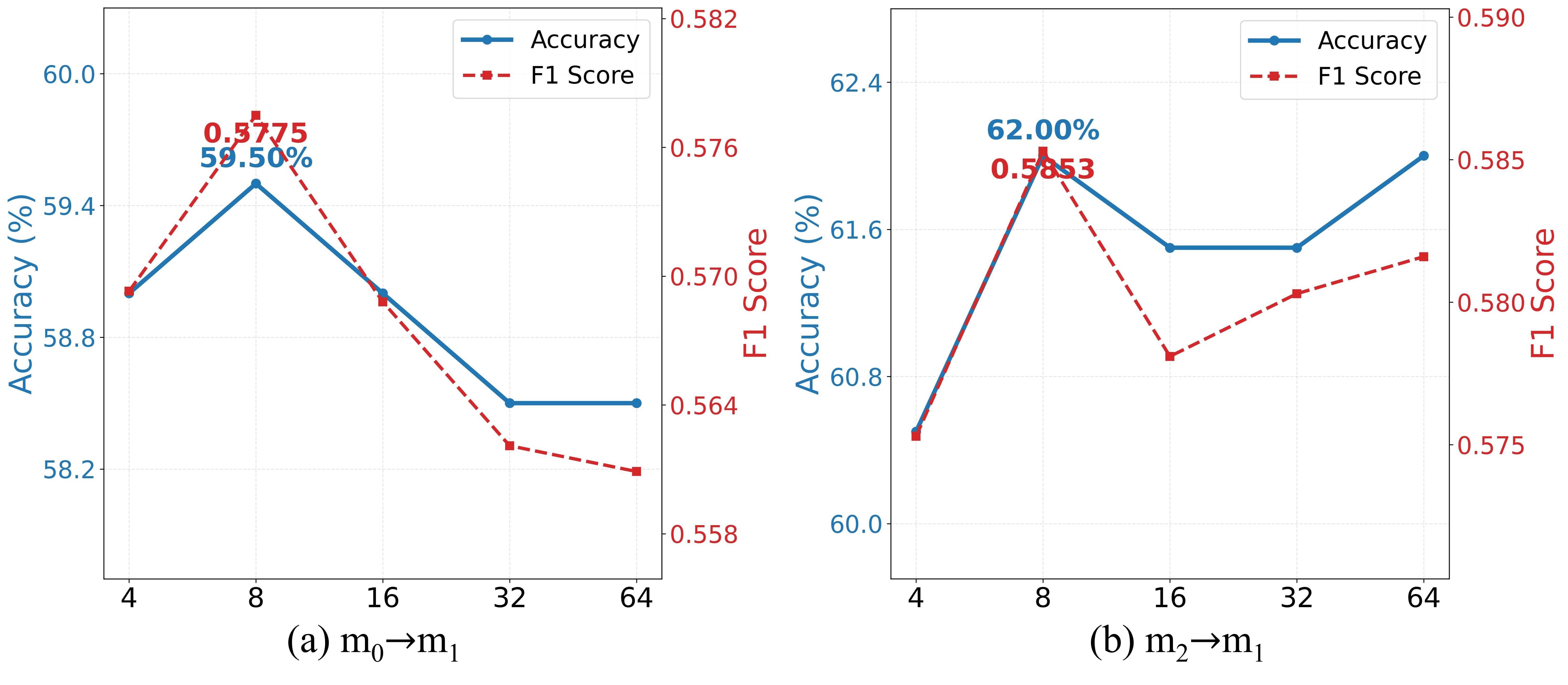}}
\vspace{-1.0em}
\caption{The sensitivity analysis of LoRA rank $r$ in two cross-mode scenarios ($m_0\to m_1$ and $m_2\to m_1$).}
\label{r_VAL}
\vspace{-1.0em}
\end{figure}

Then, we analyze the sensitivity of smoothing hyper-parameter $\sigma$ and LoRA rank $r$ in two scenarios. In Figure \ref{smooth}, $\sigma$ varies from 0.1 to 1.1. Experimental performance peaks at $\sigma=0.9$, reaching 59.50$\%$ and 62.00$\%$ with F1-scores of 0.5775 and 0.5853. As illustrated in Figure \ref{r_VAL}, the effect of LoRA rank $r$ on cross-mode radar signal recognition is evaluated for $m_0 \to m_1$ and $m_2 \to m_1$ scenarios. The proposed RadarPos attains optimal performance at $r=8$, achieving strong results with a relatively small number of trainable parameters, which highlights its efficiency.
\vspace{-8pt}

\section{Conclusions}
In this work, we propose RadarPos, a position-aware self-supervised framework for cross-mode radar recognition. By modeling inter-pulse temporal dependencies via position prediction, RadarPos enhances feature discriminability and domain alignment without heavy augmentations. Experiments across multiple scenarios show consistent gains over domain generalization and self-supervised baselines, especially in long-tailed settings. Ablation studies confirm the benefits of TOA-based encoding and the position smoothing strategy, highlighting robust generalization in dynamic environments.


\bibliographystyle{IEEEbib}
\bibliography{strings,refs}

\end{document}